\documentclass[runningheads]{svmult}
\usepackage{epsfig}
\usepackage{makeidx}   
\usepackage{graphicx}  
\usepackage{subeqnar}  
\usepackage{multicol}  
\usepackage{cropmark} 
\usepackage{physprbb}  
\def\eq{\begin{equation}}
\def\eeq{\end{equation}}
\def\eqa{\begin{equation}}
\def\eeqa{\end{equation}}
\def\roughly#1{\mathrel{\raise.3ex\hbox{$#1$\kern-.75em\lower1ex\hbox{$\sim$}}}}
\def\lsim{\roughly<}

\def\pref#1{(\ref{#1})}

\begin{document}
\title*{Natural Quintessence and the Brane World}
\toctitle{Natural Quintessence and the Brane-World}
%
%
\titlerunning{Natural Quintessence and the Brane World}
%
\author{C.P. Burgess}
\authorrunning{C.P. Burgess}
%
%
\institute{Physics Department, McGill University\\
3600 University St., Montr\'eal, Canada, H3A 2T8.}

\maketitle              

\begin{abstract}
Although quintessence models have many attractive cosmological
features, they face two major difficulties. First, it has not yet
been possible to find one which convincingly realizes the goal of
explaining present-day cosmic acceleration generically using only
attractor solutions. Second, quintessence has proven difficult to
obtain within realistic microscopic theories, largely due to two
major obstructions. Both of these difficulties are summarized in
this article, together with a recent proposal for circumventing
the second of them within a brane-world context. It is shown that
this proposal leads to a broader class of dynamics for the
quintessence field, in which its couplings slowly run (or:
`walk') over cosmological time scales. The walking of the
quintessence couplings opens up new possibilities for solving the
first problem: that of obtaining acceptable transitions between
attractor solutions.
\end{abstract}

\section{Introduction}
The discovery by cosmologists that the Universe is currently
dominated by two distinct types of unknown forms of matter is a
development with truly Copernican implications for our picture of
the Universe as a whole. We have known for some time that visible
matter likes to cluster on large scales into galaxies and galaxy
clusters, with 90\% or more of the mass of these objects
consisting of an unknown nonbaryonic `dark matter'
\cite{DMCitations}. The more recent surprise was the discovery
that this clustered dark matter itself makes up no more than 30\%
of the overall energy density of the Universe, with the remaining
70\% apparently consisting of a different kind of unknown
substance, sometimes called `dark energy' \cite{DECitations}.

It is absolutely breathtaking that so little is known about these
two most abundant forms of matter. What is known is usually
expressed in terms of their equations of state, through the ratio
of pressure to energy density, $w = p/\rho$. The formation of
galaxies and galaxy clusters by gravitational attraction appears
to require the dark matter to be `cold', with $w$ close to zero.
Similarly, the current acceleration which the universal expansion
is undergoing indicates $w \lsim -0.3$ for the dark energy.

Perhaps the simplest explanation for the dark energy is that it is
simply the energy density of the quantum vacuum, since this
satisfies $w = -1$ and is generically nonzero in realistic quantum
field theories. The problem with this explanation is that the
predicted vacuum energy is typically at least $10^{56}$ times
larger than what is observed. The Cosmological Constant Problem
\cite{CCProblem} is the recognition that at present no way is
known to naturally obtain a vacuum energy anywhere near the
required size within a realistic microscopic theory.

Quintessence models \cite{Quintessence} take a different tack to
explain the dark energy. In these models the dark energy is
attributed to the dynamics of a scalar field, $\phi$, which is
currently evolving in a cosmologically interesting way. Since $w =
(K-V)/(K+V)$, where $K= \frac12 \dot\phi^2$ and $V$ are the
scalar's kinetic and potential energies, the condition that $w$ be
negative requires the scalar's evolution at present must be slow
in the sense that $K \lsim V$. (The vacuum energy is a special
case of this kind of evolution, where $K=0$.)

These models do not solve the Cosmological Constant Problem, since
they do not provide natural reasons for $V$ to be currently so
small, but they can (potentially) explain why an evolving scalar
field could naturally have an energy density which is now so
similar to that of other forms of matter, like the dark matter.
They can do so, firstly because their equations of motion often
admit `tracking' solutions, within which the scalar energy density
closely follows (or tracks) the dominant energy density of the
Universe as it evolves. Furthermore, the late-time evolution of
the scalar field is often drawn to these `tracking' solutions for
wide choices of initial conditions, because these solutions are
also `attractors' for the scalar equations of motion.

Unfortunately, since these tracker solutions typically require the
scalar itself not to be the dominant energy density, in order to
become the dark energy the scalar must eventually leave the
tracker solution. Although one might hope that this could also be
naturally achieved -- such as being perhaps due to a transient
behaviour due to the crossover from radiation to matter domination
-- so far it has proven difficult to make a completely convincing
cosmology along these lines.

The purpose of this article is to describe a new category of
quintessence model, which could be called `Walking Quintessence'
\cite{ABRS1,ABRS2}, that may offer new ways to accomplish this
crossover. They may do so because within these models the
couplings of the scalar field slowly run (or walk) as the Universe
evolves, and this walking may help facilitate the crossover between
tracking solutions. What is remarkable is that these models were
developed in an attempt to address a completely different set of
(very serious) problems which arise once one tries to obtain
quintessence from a realistic model of microscopic physics.

Since they play such an important part in the motivation of the
models, the bulk of the article is devoted to these serious
microphysical problems. The problems themselves are first
summarized in the next section, followed by a description how they
are addressed within the attractive brane-world picture which has
emerged as a potential low-energy consequence of string theory.
The results of a sample cosmology built on this model are then
briefly presented, intended as first step towards a more
systematic exploration of the cosmologies which are suggested by
this class of models.

\section{Naturalness Issues}
{}From a microscopic perspective any viable quintessence model
must have two very remarkable properties, which turn out to be
quite difficult to arrange \cite{Carroll}. 
As is explained below, they must have:
\begin{itemize}
\item
{\bf Extremely Light Scalars}, whose mass must be at most $m_Q =
10^{-32}$ eV, and;
\vspace{3 mm}
\item
{\bf No New Long-Range Forces}, to the extent that these would
ruin the agreement between General Relativity and observations on
Earth, within the Solar System and beyond \cite{Will,ABFN}.
\end{itemize}

\subsection{Light Scalars}
The first requirement (very light scalars) is a very generic
property of explanations of the dark energy in terms of a rolling
scalar field. Its necessity may be seen from the scalar field
equation of motion within a cosmological context:
\eq \label{KGEq} \ddot{\phi} + 3 \, H \, \dot\phi + {\partial V
\over \partial \phi} = 0, \qquad \hbox{and} \qquad H^2 = {\rho
\over 3 M_p^2}, \eeq
where $M_p = 10^{18}$ GeV is the rationalized Planck mass, and
$\rho$ is the total energy density (which is at present dominated
by the dark energy).

To quantify how small this requires the scalar mass to be, it is
instructive to consider the very broad class of models within
which
\eq \label{Vform} V(\phi) = \mu^4 \; U(\phi/f) . \eeq
Here $\mu$ and $f$ are arbitrary mass scales, whose size may be
determined if the dimensionless function $U(x)$ and its
derivatives are at present ${\cal O}(1)$. In this case the present
value of the scalar potential and its derivatives are $V\sim
\mu^4$, $dV/d\phi \sim \mu^4/f$, {\it etc.}, and the square of the
scalar mass is of order $d^2 V/d\phi^2 \sim \mu^4/f^2$. This turns
out to be of order $(10^{-33}\, \hbox{eV})^2$ given the values
which are required for $\mu$ and $f$.

The value $\mu \sim 10^{-3}$ eV is inferred by recognizing that
these estimates also apply to the total scalar field energy, $\rho
= K + V \sim V$, since present observations require the scalar
field to be rolling with $K$ less than but of the same order of
magnitude as $V$. In this way we see that $\mu$ controls the
present dark energy density, $\rho \sim \mu^4 \sim (10^{-3} \;
\hbox{eV})^4$, and so also $H \sim \mu^2/M_p$ and $\dot\phi \sim
\sqrt{K} \sim \sqrt{V} \sim \mu^2$.

The value $f = M_p$ is determined from the scalar field equation,
eq.~\pref{KGEq}, together with the above slow-roll conditions,
since this implies the $\ddot\phi$ term should be much smaller
than the other two. Except for the case of a pure cosmological
constant (for which only the last term is important), we therefore
have $H \dot\phi \sim\partial V/\partial \phi$ and so $\mu^4/M_p
\sim \mu^4/f$, from which we learn $f\sim M_p$ and hence $m_Q \sim
\mu^2/f \sim 10^{-33}$ eV.

Such an extraordinarily small scalar mass is extremely difficult
to achieve in a realistic microscopic theory. There are two
separate aspects to this difficulty.
\begin{enumerate}
\item
{\bf Heirarchy Problem 1:} How does such a small nonzero mass
arise as a combination of microscopic parameters?
\vspace{3 mm}
\item
{\bf Heirarchy Problem 2:} Given that such a small mass is
predicted by the theory of microscopic physics, how does it remain
small as one integrates out all the physics between these
microscopic scales and the cosmological scales at which it is
measured? This is a problem because, for instance, a particle of
mass $M$ and coupling $1/f$ shifts the scalar mass by an amount
\eq \delta m \sim {M^2 \over 4 \pi f} \eeq
when it is integrated out.
\end{enumerate}

Both of these problems are the direct analogs of two aspects of
the famous heirarchy problem as applied to the Higgs field which
breaks electroweak gauge symmetry within the Standard Model of
particle physics. There the weak scale, $M_w \sim 100$ GeV is
controlled by the scalar Higgs mass, and one asks how this can be
so much smaller than, say, the Planck scale, $M_p \sim 10^{18}$
GeV.

The problem for quintessence models, however, is arguably much
worse for two reasons. First, the quintessence scalar is many more
orders of magnitude lighter than the basic microscopic physics
scale $M_w$ than $M_w$ itself is from $M_p$. Second, the range of
scales between $M_w$ and $m_Q$ is well studied by experiment,
which presumably makes it harder to hide the other degrees of
freedom which are typically invoked to alleviate Heirarchy Problem
2 (such as the superpartners which do the job if supersymmetry is
the solution).

Almost none of the extant quintessence models address these
issues, with the exception being those based on pseudo-Goldstone
bosons \cite{PSGB1,PSGB1a}, which address Heirarchy Problem 2.
(Modifications of these models based on the brane world can also
address Heirarchy Problem 1 \cite{PSGB2}.) The models presented in
the later sections are unique among those yet proposed in that
{\emph both} of these problems are related to the same microscopic
quantities which explain why $M_w \ll M_p$.

\subsection{Long-Range Forces}
The incredibly small quintessence scalar mass also raises a
related observational problem, since it implies that the exchange
of this scalar must mediate a very long-range force. Furthermore,
the strength with which this force couples to particles of energy
$E$ is typically of order $E/f \sim E/M_p$, which makes it comparable to
gravity. This is problematic, since many observations now strongly
constrain the existence of gravitational-strength forces between
macroscopic objects having a
range longer than about 0.1 mm.

In the models which are described in subsequent sections this
problem is evaded because the scalar couplings turn out to evolve
over cosmological time scales. In the examples given it happens
that these couplings evolve to become extremely small during the
present epoch, which is when all of the very constraining
observations have been made.

Another way to ensure acceptably small couplings is to arrange the
scalar to couple to quantities, such as spin, which ordinary
matter in bulk does not carry. This kind of coupling can occur in
pseudo-Goldstone-boson models \cite{PSGB1,PSGB2}.

\section{Large Extra Dimensions}
New insights into naturalness problems, like the heirarchy between
$M_w$ and $M_p$, have been made based on the recently
much-discussed brane world picture, in which all observed
elementary particles are confined to a domain-wall-like surface
(or `brane') which sits within a higher-dimensional `bulk'
spacetime. The simplest choice puts us on a 4-dimensional surface
(or 3-brane) within a bulk space which has anywhere from 5 to 11
dimensions. By contrast, gravitational interactions in this
picture are not confined to these branes. This kind of picture is
very well motivated within string theory.

The fact that gravity and other interactions do not see the
same number of dimensions lies at the root of the surprising
realization that the fundamental string scale, $M_s$, can be much
smaller than $M_p$ \cite{LowSS}, and of the new perspective this
has allowed for understanding low-energy naturalness problems. In
particular, it can imply the relationship
\eq \label{WPexp} \left( {M_s \over M_p} \right)^2 \sim {\alpha^2
\over (M_s r)^n}, \eeq
where $\alpha$ is the (open) string coupling and $r$ is the radius
of the $n$ extra dimensions. For supersymmetric systems one
generically has $M_w/M_p \sim (M_s/M_p)^2$ and so this remarkable
formula allows the observed ratio $M_w/ M_p \sim 10^{-16}$
(Hierarchy Problem 1, above) to be understood using parameters
themselves no smaller than $\alpha \sim 1/(M_s r) \sim 0.01$ if
$M_s\sim 10^{11}$ GeV and there are $n=6$ extra dimensions
\cite{BBIQ}. (In this picture it is supersymmetry which accounts
for Heirarchy Problem 2.)

\subsection{Natural Quintessence and the Brane World}
The brane-world variant which is of most interest for the present
purposes, puts the string scale as low as is consistent with
experiment, $M_s \sim M_w$ \cite{ADD}. From eq.~(\ref{WPexp}), one
sees this is permitted (even if $\alpha$ is not small) provided
$r$ is large enough. For instance, if 
$n=2$ then $r \sim 0.1$ mm (or $1/r \sim 10^{-3}$ eV). In this
picture $M_w/M_p \sim \alpha/(M_s r)$, so the heirarchy problem is
not solved so much as translated into the problem of understanding
the origin of the large heirarchy $M_s r/\alpha \sim 10^{16}$.

A remarkable feature of this scenario is that it provides a
framework within which scalars can be naturally as light as
$10^{-33}$ eV, and it is instructive to see how this works for
specific examples. The success of the models of
ref.~\cite{ABRS1,ABRS2} relies on choosing the extra dimensions to
be torii and the quintessence field to be a component of the
extra-dimensional metric -- the radion, $r$.

This construction directly solves Heirarchy Problems 1 as follows.
First, because the radion field starts life as a component of the
six-dimensional metric its kinetic term has the same origin as
does the 4D graviton. Dimensionally reducing the 6D
Einstein-Hilbert action to four dimensions gives
\eq \label{4DKin} {{\cal L}_{\rm kin} \over \sqrt{- g}} = - \,
{M_p^2 \over 2} \; g^{\mu\nu} \left[ R_{\mu\nu} + {4 \over r^2}
\partial_\mu r \partial_\nu r \right] , \eeq
from which we see the canonically-normalized field is $\phi$
where $r = r_0 \exp(\phi/2M_p)$. From this it follows that $f \sim M_p$.

For toroidal compactifications, direct dimensional reduction of
the higher-dimensional Einstein-Hilbert action gives no radion
potential at all (provided that the cosmological constant is
chosen to vanish, as usual). This is an artifact of the classical
approximation, however, and a potential for $r$ is generated once
quantum effects are included, such as through the Casimir effect
which predicts (for large $r$) a potential of the form $V \sim
1/r^4$ \cite{ABRS1,PP}. If this potential can be stabilized to
have a minimum for some $r = r_0$ (more about this later), then it
predicts $\mu \sim 1/r_0$.

More remarkably, this model also addresses Heirarchy Problem 2,
since these predictions are protected from being ruined as physics
between the weak scale and the quintessence mass is integrated
out. Although the stability of the prediction for $f$ against
quantum corrections is fairly trivial, it is the stability of the
potential which bears closer examination.

There turn out to be two reasons for this success. The Casimir
effect predicts $\mu \sim 1/r$ largely because the potential is
generated when modes having energies of order $1/r$ are integrated
out. Now consider the effect of integrating out modes with
energies lower than $1/r$. For scales $M \lsim 1/r$ the effective
theory is four-dimensional, and the naive corrections to $V(r)$
are correct, since no symmetries preclude generating a radion
potential. Integrating out a mode of mass $M$ then contributes to
$V(r)$ terms of order $\delta \mu^4 \sim M^4/(4 \pi)^2$, which is
not dangerous since it predicts $\delta \mu \sim M/\sqrt{4 \pi}
\lsim 1/r$.

The key is at energies above the scale $1/r$, where the effective
theory is six-dimensional and so is constrained by additional
symmetries like 6D general covariance. For these scales the result
of integrating out modes with $M \gg 1/r$ may be expanded in
powers of the 6D curvature $R_{mnpq}$. If the extra dimensions
were to have spherical geometry then $R_{mnpq} \propto 1/r^2$, and
these curvature terms constitute dangerously large contributions
to the radion potential. For flat spaces like torii, however,
$R_{mnpq}$ is independent of $r$ and these terms are not
dangerous. The question for torii becomes whether it is possible
to keep the internal dimensions flat despite the existence of
large vacuum energies on the various branes on which our observed
particles live. It is a special feature of two dimensions that
this is so, since Einstein's equations predict flat geometries
around point sources \cite{2DFlat}.

We see that in this picture the quintessence scales $\mu$ and $f$
are predicted to be respectively given by $1/r$ and $M_p$ because
the quintessence field has its microscopic origin as part of the
higher-dimensional geometry. The success of the predicted dark
energy density, $V \sim 1/r^4$, and quintessence mass, $m_Q \sim
1/(M_p r^2)$, then follow from the choice $1/r \sim 10^{-3}$ eV
which is required in any case to solve the electroweak heirarchy,
since $M_w/M_p \sim 1/(M_w r)$. This success does not crucially
hinge on identifying the quintessence field as a mode of the
metric, since variant brane-world models with similar properties
may also be built wherein the quintessence field is a
pseudo-Goldstone mode in the bulk \cite{PSGB2}.

\section{Radius Stabilization and `Walking' Quintessence}
In order to more precisely pin down the cosmology the explicit
form for the quintessence potential is required, and within the
brane world picture being presented this amounts to providing a
mechanism for stabilizing the radion at large values. This section
relates a concrete proposal for doing so, as made in
ref.~\cite{ABRS1}.

Besides providing an interesting cosmology in its own right, there
is another reason for exploring this specific proposal in more
detail. This other reason is to illustrate why worrying about
naturalness issues is useful when exploring phenomenological
applications (such as quintessence cosmology). Naturalness issues
are useful precisely because the cosmology of quintessence models
depends in such a detailed way on the precise form of the
low-energy potential. On one hand, the naturalness problem states
that only very specific kinds of potentials are likely to arise as
the low-energy limit of realistic microphysics. On the other hand,
it has proven difficult to build a completely convincing
quintessence cosmology purely by guessing different kinds of
potentials. It may be true that it is only the very few potentials
which can arise from real microphysics that can also provide a
realistic description of cosmology as well. If so, it should be
invaluable to be able to explore in detail any potential which
{\emph does} arise as the low-energy limit of real microphysics.

In the present instance the stabilization mechanism proposed
suggests a qualitative new feature of the low-energy theory: it
predicts that the effective low-energy couplings and masses depend
logarithmically on the quintessence field, and so these all slowly
run ({\it i.e.} walk) over cosmological time scales. This leads in
a natural way to extended quintessence models \cite{ExtQuint}, but
with a specific and cosmologically-interesting type of
field-dependence. Indeed the proposal of ref.~\cite{ABRS1} was
initially motivated by the earlier discovery of the attractive
cosmologies which can result from quintessence models having
exponential potentials with logarithmic corrections \cite{AS}.

\subsection{Stabilization via Six-Dimensional Logarithms}
The models now described are based on the observation that $V(r)$
would naturally be minimized at large values for $r$ if it were to
depend logarithmically on $r$:
\eq \label{LogForm} V(r) = V_0 \left({\ell \over r}\right)^p
\left[1 + \epsilon \log\left({r \over \ell} \right) \right] +
\dots , \eeq
where $a$ and $\epsilon$ are constants and $\ell$ is a microscopic
length scale, such as $\ell \sim 1/M_s$. The ellipses indicate
other terms which fall off with a higher power of $\ell/r$
relative to those shown. A potential of this form has a minimum at
$r_0 \sim \ell \, \exp(1/\epsilon)$, which is exponentially larger
than $\ell$ if $\epsilon$ is moderately smaller than one. (Values
$\epsilon \sim 1/50$ are sufficient to generate the desired
heirarchy in the models explored in \cite{ABRS1}.)

The key point is that a potential of precisely this form is
expected if the effective six-dimensional theory at scales $E >
1/r$ contains a renormalizable ({\it i.e.} marginal) coupling,
$g$. If so, then this coupling runs logarithmically with $r$ and
loop corrections to the radion potential involving only this
coupling have the logarithmic form of eq.~\pref{LogForm}, with a
coefficient $\epsilon \sim g^2/(4 \pi)^3$ which is naturally
small. In six dimensions most interactions are not renormalizable,
but such couplings can exist, such as a cubic coupling amongst
six-dimensional scalar fields.

Given this generic mechanism it is clear that logarithms in $r$
are not restricted to appear in the low-energy theory only within
the radion potential. Rather, they arise generically as radiative
corrections to {\emph all} couplings, and it is this logarithmic
dependence which is responsible for the walking of these couplings
as $r$ evolves over cosmological time scales.

\subsection{A Sample Quintessence Cosmology}
Logarithmically evolving couplings are very attractive for
quintessence cosmology for two reasons. First, they introduce
non-minimal couplings between the quintessence field and ordinary
matter, and these turn out to provide new kinds of scaling
attractor solutions. Which attractor is the endpoint for a given
choice of initial conditions is determined by the parameters of
the quintessence potential, and this observation is at the root of
the second attractive cosmological feature of having walking
couplings. A given cosmological solution can cross over from the
basin of attraction of one attractor to another since the walking
of the couplings moves the boundaries of these domains of
attraction within field space.

Ref.~\cite{ABRS2} provides a preliminary exploration of the
cosmology of models built along these lines. An example of a
viable cosmology which is obtained in this way is illustrated in
Figure 1, which shows the cosmological evolution of the energy
density in radiation, matter and the radion kinetic and potential
energies. Figure 2 shows, for the same model, the coupling,
$\eta$, of the radion to ordinary matter as a function of the
universal scale factor, from which it is clear that the model
evades current constraints on long-range forces (which require
$\eta < 0.03$).

A constraint which is specific to models where it is the radion which
is the quintessence field is the requirement that the radion not
roll too far since the epoch of nucleosynthesis, since such a roll
would predict an unacceptably large change in the ratio of weak
and gravitational couplings, $M_w/M_p \sim 1/(M_s r)$, between
then and now. Figure 3 shows how $M_s r$ evolves for the cosmology
shown in Figure 1, and so demonstrates that its evolution can be
acceptably small.

Although a promising start, this cosmology still leaves much
to be desired. In particular, the constancy of $r$ is not
generic and must be partially
arranged by adjusting the parameters of the potential 
and the initial energy in the scalar roll. However the
broader framework of walking quintessence models seems to
provide a fruitful place to search for a quintessence model
which is truly natural in all senses of the word.

\begin{figure}[1]
\begin{center}
\includegraphics[width=.5\textwidth]{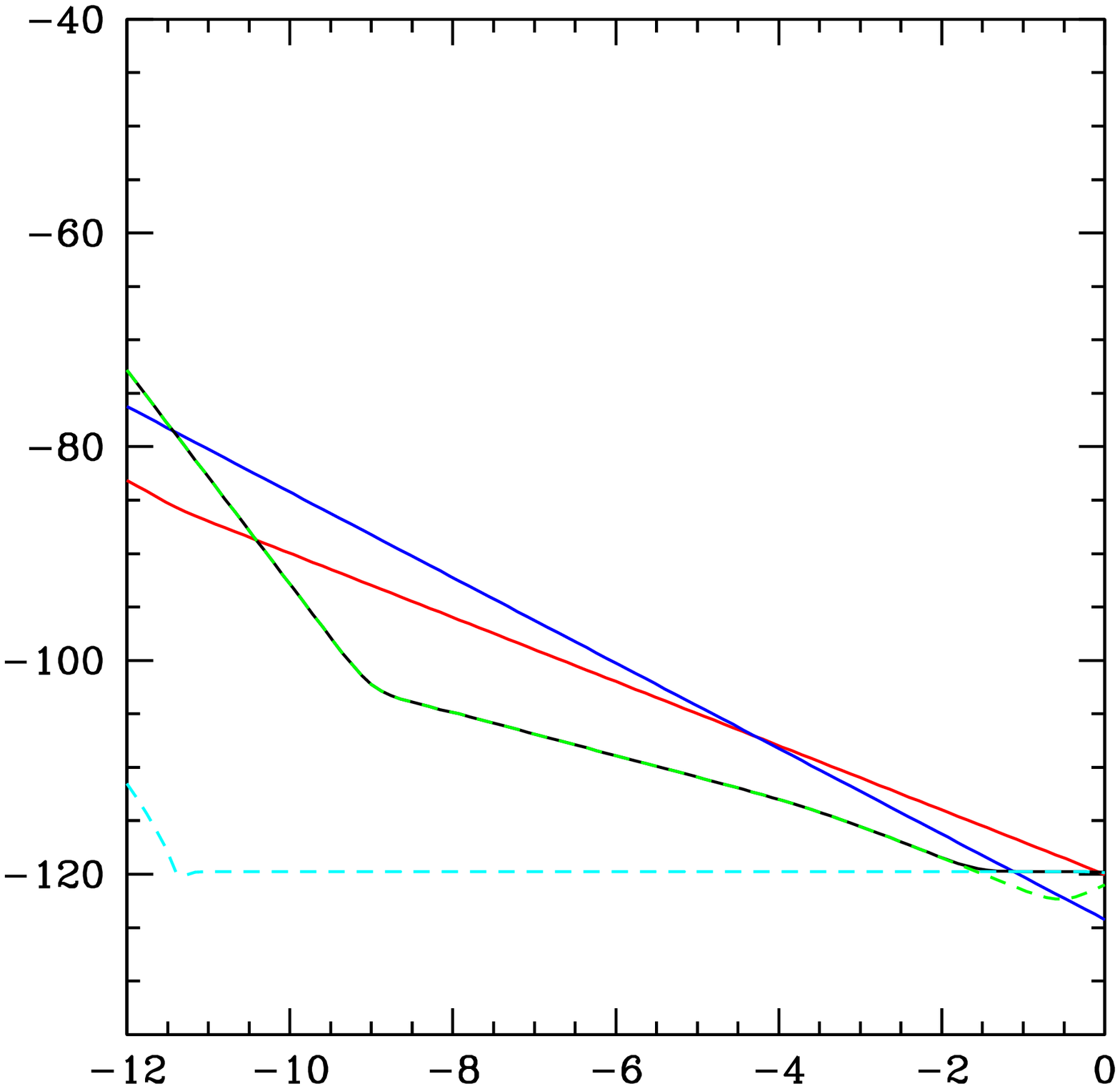}
\end{center}
\caption[]{The logarithm of the energy density of radiation (solid
blue), matter (solid red), scalar kinetic (green dashed) and
radion potential (cyan dashed), plotted against the logarithm of
the universal scale factor (normalized to unity at present).}
\label{fig1}
\end{figure}

\begin{figure}[2]
\begin{center}
\includegraphics[width=.5\textwidth]{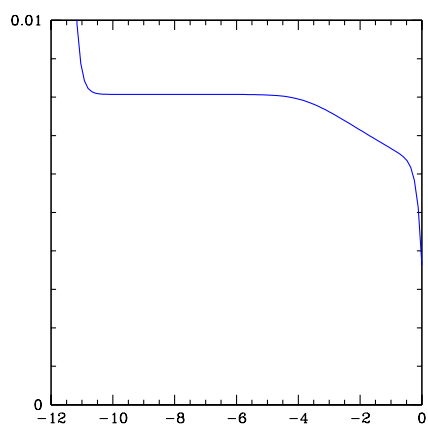}
\end{center}
\caption[]{The quintessence-matter coupling, $\eta$, plotted
against the logarithm of the scale factor for the same cosmology
as Figure 1. The observational constraint is $\eta < 0.03$ during
the present epoch.} \label{fig2}
\end{figure}

\begin{figure}[3]
\begin{center}
\includegraphics[width=.5\textwidth]{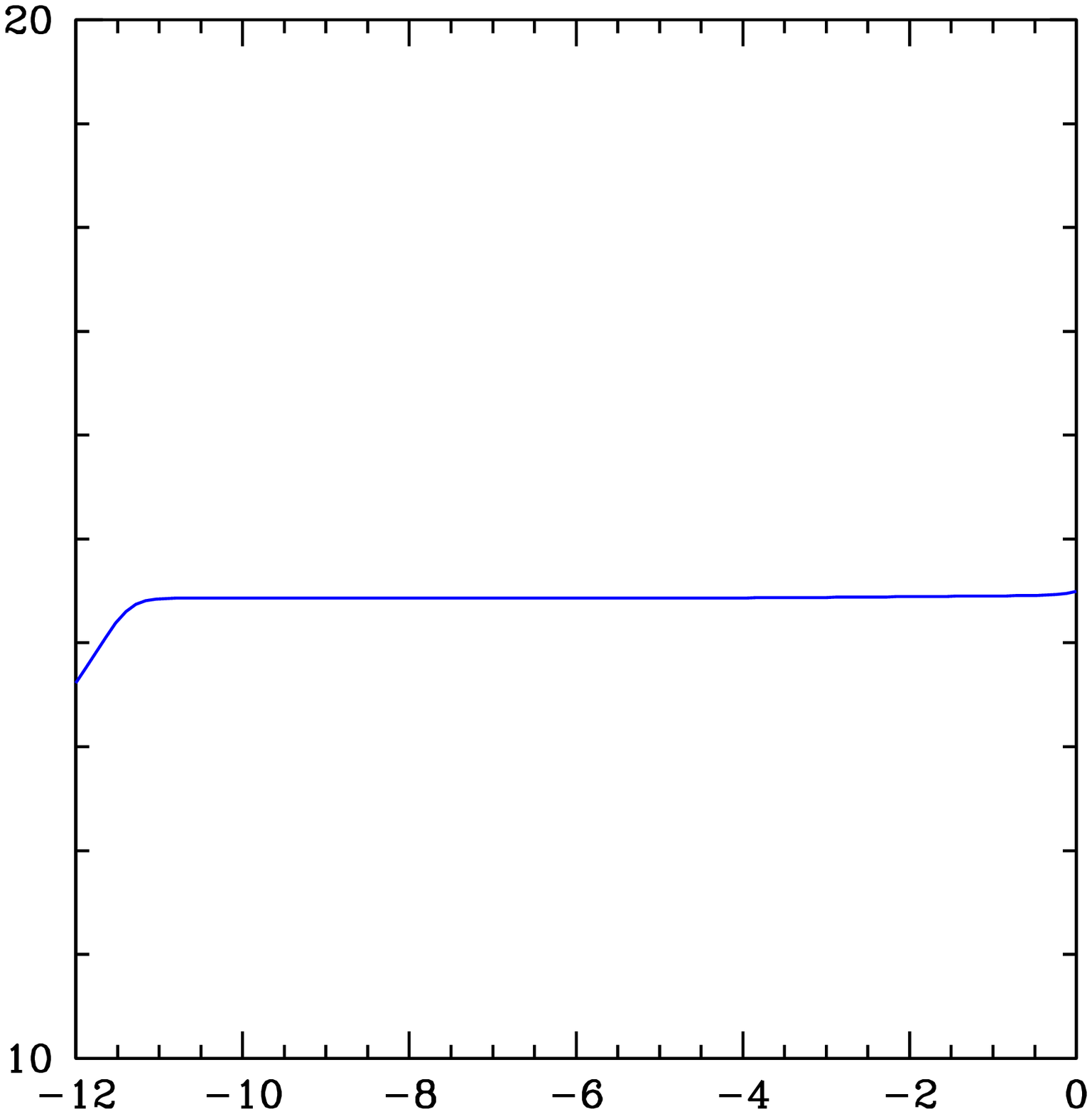}
\end{center}
\caption[]{The logarithm of the ratio $M_p/M_w \sim M_s r$ against
scale factor for the same cosmology as for Figure 1. This ratio
must be within roughly 10\% of its present value $10^{16}$ at
nucleosynthesis ($a \sim 10^{10}$). }
\label{fig3}
\end{figure}


\vspace{1 cm}
\begin{center}
{\bf Acknowledgements}
\end{center}
I would like to thank the organizers of this workshop for
providing such stimulating environs, and for their kind invitation
to speak. My part of the work summarized in this article was done
in collaboration with Andy Albrecht, Finn Ravndal and Constantinos
Skordis. It was started at the Aspen Center for Physics, and was
largely carried out at the Institute for Advanced Study in
Princeton with financial help from the Ambrose Monell Foundation.
My own research during this period was supported by grants from
N.S.E.R.C. (Canada) and F.C.A.R. (Qu\'ebec).


\begin{thebibliography}{8.}
\addcontentsline{toc}{section}{References}

\bibitem{DMCitations} 
P.J.E. Peebles, {\it Publ. Astron. Soc. Pacific} {\bf 111} (1999) 274;
M.S. Turner, {\it Publ. Astron. Soc. Pacific} {\bf 111} 264;
A. Dekel, D. Burstein and S.D.M. White, {\it Critical Dialogues in Cosmology} (1997) p. 175.

\bibitem{DECitations} S.~Perlmutter et al., Ap. J. {\bf 483} 565 (1997) (astro-ph/9712212);
A.G. Riess {\it et al}, Ast. J. {\bf 116} 1009 (1997)
(astro-ph/9805201); N. Bahcall, {\it et al.} {\it Science} {\bf
284}, 1481, (1999).

\bibitem{CCProblem} S. Weinberg, {\it Rev. Mod. Phys.} {\bf 61} (1989) 1--23.

\bibitem{Quintessence} 
K.~Coble, S.~Dodelson and J.~A.~Frieman,
Phys.\ Rev.\ D {\bf 55}, 1851 (1997)
(astro-ph/9608122);
R.~R.~Caldwell and P.~J.~Steinhardt,
Phys.\ Rev.\ D {\bf 57}, 6057 (1998)
(astro-ph/9710062);
I.~Zlatev, L.~M.~Wang and P.~J.~Steinhardt,
Phys.\ Rev.\ Lett.\  {\bf 82}, 896 (1999)
(astro-ph/9807002);
P.~J.~Steinhardt, L.~M.~Wang and I.~Zlatev,
Phys.\ Rev.\ D {\bf 59}, 123504 (1999)
(astro-ph/9812313).

\bibitem{ABRS1}
A. Albrecht, C.P. Burgess, F. Ravndal and C. Skordis, {\it Phys.
Rev.} {\bf D} (to appear) (hep-th/0105261).

\bibitem{ABRS2}
A. Albrecht, C.P. Burgess, F. Ravndal and C. Skordis, {\it Phys.
Rev.} {\bf D} (to appear) (astro-ph/0107573).

\bibitem{Carroll}
For a recent summary of these issues see, 
for example, S. Carroll, (astro-ph/0107571).

\bibitem{Will}
C.~M.~Will, (gr-qc/9811036);
C.~M.~Will,
Living Rev.\ Rel.\  {\bf 4}, 4 (2001)
(gr-qc/0103036).

\bibitem{ABFN}
A. Aguirre, C.P. Burgess, A. Friedland and D. Nolte, {\it Class.
Quant. Grav.} {\bf 18} (2001) R223--R232, (hep-ph/0105083).

\bibitem{PSGB1}
J.~A.~Frieman, C.~T.~Hill, A.~Stebbins and I.~Waga,
Phys.\ Rev.\ Lett.\  {\bf 75}, 2077 (1995)
(astro-ph/9505060).

\bibitem{PSGB1a}
For more recent work see
C.T. Hill, A.K. Leibovich, FERMILAB-PUB-02-085-T, May 2002, 
(hep-ph/0205237).
 
\bibitem{PSGB2}
C.P. Burgess {\it et.al.}, in preparation.

\bibitem{LowSS}
P.~Horava and E.~Witten, Nucl.\ Phys.\ {\bf B475} (1996) 94--114
({\tt hep-th/9603142}); Nucl.\ Phys.\ {\bf B460} (1996) 506--524
({\tt hep-th/9510209}); E.~Witten, Nucl.\ Phys.\ {\bf B471} (1996)
135 ({\tt hep-th/9602070}); J. Lykken, {\it Phys. Rev.} {\bf D54}
(1996) 3693 -- 3697 ({\tt hep-th/9603133}); I.~Antoniadis, Phys.\
Lett.\ {\bf B246} (1990) 377 -- 384.

\bibitem{BBIQ}
C.P.~Burgess, L.E.~Iba\~nez and F.~Quevedo, {\it Phys. Lett.} {\bf
B447} (1999) 257 (hep-ph/9810535); K.~Benakli, {\it Phys. Rev.}
{\bf D60} (1999) 104002 (hep-ph/9809502).

\bibitem{ADD}
N. Arkani-Hamed, S. Dimopoulos and G. Dvali, {\it Phys. Lett.}
{\bf B429} (1998) 263--272 ({\tt hep-ph/9803315}); Phys.\ Rev.\
{\bf D59} (1999) 086004 ({\tt hep-ph/9807344}); I.~Antoniadis,
N.~Arkani-Hamed, S.~Dimopoulos and G.~Dvali, Phys.\ Lett.\ {\bf
B436} (1998) 257--263 ({\tt hep-ph/9804398}).

\bibitem{PP}
E. Pont\'on and E. Poppitz, {\it JHEP} 0106 (2001) 019 (hep-th/0105021).

\bibitem{2DFlat}
S. Deser, {\it Annals Phys.} {\bf 152} (1984) 220; A. Vilenkin and T. Vachaspati, 
{\it Phys. Rev.} {\bf 35} (1987) 1138.

\bibitem{ExtQuint}
F. Perrotta, C. Baccigalupi and S. Matarrese, {\it Phys. Rev.} {\bf D61} (2000) 023507 (astro-ph/9906066).

\bibitem{AS}
A. Albrecht and C. Skordis, {\it Phys. Rev. Lett.} {\bf 84} 2076
(2000) (astro-ph/9908085).


\end{thebibliography}
\end{document}